\documentstyle[aps,prb]{revtex}
\begin{document}
\draft
\title{Composition Dependence of the Structure and \\
Electronic Properties of Liquid Ga-Se Alloys Studied by \\
Ab Initio Molecular Dynamics Simulation}
\author{J.~M.~Holender \cite{jh} and M.~J.~Gillan \cite{mg}}
\address{Physics Department, Keele University, Staffordshire ST5 5BG, U.K.}
\maketitle

\begin{abstract}
{\em Ab initio} molecular dynamics simulation is used to study the
structure and electronic properties of the liquid Ga-Se system
at the three compositions Ga$_2$Se, GaSe and Ga$_2$Se$_3$, and of
the GaSe and Ga$_2$Se$_3$ crystals. The calculated equilibrium structure
of GaSe crystal agrees well with available experimental data. The
neutron-weighted liquid structure factors calculated from the
simulations are in reasonable agreement with recent neutron diffraction
measurements. Simulation results for the partial radial distribution
functions show that the liquid structure is closely related to
that of the crystals. A close similarity between solid and liquid
is also found for the electronic density of states and charge density.
The calculated electronic conductivity decreases strongly with increasing
Se content, in accord with experimental measurements.
\end{abstract}

\section{Introduction}
The study of liquid semiconductor alloys goes back many years,
and has been a splendid source of insights into the relationships
between atomic ordering, chemical bonding and electrical properties
in condensed matter~\cite{enderby90}.
In some well studied systems like
Cs-Au~\cite{schmutzler76} and Mg-Bi~\cite{enderby70},
where the elements have rather different electronegativities,
the electrical conductivity passes from highly metallic values to insulating
values, as a gap at the Fermi level opens up in the density of states and
the bonding changes from metallic to ionic at the stoichiometric
composition. In cases where one element
is a metal and the other a semiconductor in the liquid state, the chemical
bonding can be continuously tuned from metallic to covalent as the
composition is varied. The Ga-Se system studied in this paper is
an example of the latter kind.

We have used {\em ab initio} molecular dynamics
simulation~\cite{cp} (AIMD)
to study the liquid Ga-Se alloy at the three compositions
Ga$_2$Se, GaSe and Ga$_2$Se$_3$. The simulations have been used
to study the liquid structure, the electronic density of states and the
spatial distribution of electrons as a function of composition. We shall
show that the relationship between the properties of the solid and
the liquid is very instructive, and we have used the same methods to
study the two stoichiometric solids GaSe and Ga$_2$Se$_3$.

The liquid Ga-Se system has been rather little studied experimentally,
and we chose to work on it mainly because we were aware of
plans to perform neutron diffraction
measurements on the system~\cite{gaseexp}.
These measurements, as well as studies
of the electrical conductivity, were completed while the
work was in progress, and we shall present comparisons with
both sets of results.
The system is also expected to show quite a close
resemblance to the liquid alloys In-Se (the electronic properties have
been studied by Okada and Ohno \cite{okada93}), Ga-Te (structure
measurements have been done by Takeda {\em et al.}~\cite{takeda83}
and
Hoyer {\em et al.~}\cite{hoyer84}
and electronic properties by Valiant and Faber
\cite{val74})
and In-Te (structure measurements have been made by  Hoyer {\em et al.~}
\cite{hoyer82}, conductivity and
thermopower by Popp {\em et al.~} \cite{popp74}).

It is useful to consider what structural and electronic changes would
be expected as the composition of $\ell$-Ga-Se is varied. Liquid Ga is
a typical liquid metal, with a coordination number of $\sim$~9
\cite{bel89}, while
$\ell$-Se has a covalently bonded chain-like structure with a
coordination number of 2 (Ref. \onlinecite{inui90}).
It is clear then that there must be radical
changes of structure across the composition range. However, the
electronegativity difference between Ga and Se
is fairly small, so that strongly
ionic bonding is not expected. This is confirmed by the structure
of the crystals. In the rather unusual layer
structure of GaSe~\cite{kuhn75},
the Ga atoms are tetrahedrally bonded to Se and to each other,
and the Se atoms have threefold coordination to Ga. The Ga$_2$Se$_3$
crystal~\cite{mikkelsen81} shows tetrahedral
coordination of both elements. In the middle
concentration range of the liquid,
we might therefore expect a covalently bonded
low-coordinated structure dominated by Ga-Ga and Ga-Se bonds. This
expectation is fully confirmed by our simulations.

The AIMD technique is ideally suited to this type of problem. The basic
principle is that the total energy of the system and the forces on
the atoms for any atomic arrangement are calculated by solving
Schr\"{o}dinger's equation for the valence electrons to find the
self-consistent ground state. By simulating the system in thermal
equilibrium, we therefore generate the liquid structure, the chemical
bonding and the electronic structure in a completely unified way, and
without any prior assumptions or adjustable parameters. In the
past few years, AIMD has become a widely used technique for studying
liquid metals and semiconductors~\cite{stich89,liqab}. The work reported here
is closely related to our recent simulations of $\ell$-Ga \cite{ga95}
and $\ell$-Ag-Se \cite{agse95}.

The paper is organized as follows. In the following section, we summarize
the AIMD techniques we use. Sec. 3 presents our calculations on the
equilibrium structure and electronic structure of the crystals. Our AIMD
simulations on the liquid  alloys are then reported in Sec. 4. The
significance of the results and comparisons with related liquid
alloys are discussed in Sec. 5.

\section{Techniques}
The first-principles simulation methods used here are the same
as those used in our work on $\ell$-Ga \cite{ga95},
and we give only a brief summary
of the main points. Only valence electrons are treated explicitly,
and the core states are assumed to be identical to those in the
free atoms. For the Ga-Se system, all states up to and including the 3$d$
states in both species are treated as part of the core. The interactions
between valence electrons and the cores are represented by norm-conserving
nonlocal pseudopotentials, which are constructed {\em ab initio}
via calculations on the free atoms (see below). The calculations are
performed in periodic boundary conditions, with the electronic orbitals
expanded in plane waves, all plane waves being included whose kinetic
energy is less than a chosen cutoff $E_{\rm cut}$. The exchange-correlation
energy is represented by the local density approximation (LDA), the
form used here being that due to Ceperley and Alder~\cite{lda}.
In the dynamical simulations, the ions follow classical trajectories
determined by the Hellmann-Feynman forces obtained from the
first-principles calculations, while the electronic sub-system
remains in the ground state at every instant (the Born-Oppenheimer
principle).

The problems of performing accurate first-principles simulations on
metallic systems have been extensively discussed in the literature
\cite{ga95,gil89,gru94,kre94}.
We use here the Fermi-smearing technique with orbital occupation
numbers treated as dynamical variables. The smearing is handled with
the quasi-Gaussian smearing function introduced in our $\ell$-Ga work
\cite{ga95}.
Minimization of the (free) energy to obtain the self-consistent
ground state for each ionic configuration is performed by the
preconditioned conjugate gradients method. The calculations were
performed with the all-bands version of the CASTEP code~\cite{payne92}
running on the Fujitsu VPX~240 at Manchester.

The norm-conserving pseudopotential for Ga is identical to the one used
in our $\ell$-Ga work. For Se, we generated a
pseudopotential using the standard Kerker~\cite{ker} method. The
$s$ and $p$ components of the pseudopotential were generated using
the atomic configuration 4$s^2$4$p^4$, and the $d$ component using
the configuration 4$s^2$4$p^{2.75}$4$d^{0.25}$. The core radii were taken
to be 2.0, 2.0 and 2.3~a.u. for $s$, $p$ and $d$ components respectively.
The pseudopotentials were represented in the Kleinman-Bylander form
\cite{kb},
with the $p$-wave treated as local for both Ga and Se, and the non-local
parts of the pseudopotentials treated in real space \cite{kin91}.

The electronic densities of states (DOS) for the solids were calculated
using the standard tetrahedron method~\cite{jep71}.
For the liquids we used two approaches. The first method was to use
$\Gamma$-point sampling only and to average over all ionic configurations,
and the second was to use many $k$-points for a few chosen
configurations. Both methods gave essential the same DOS.
\section{Crystalline phases}
The $\beta$-GaSe crystal has the layered structure shown in Fig.
\ref{b-gase},
with every layer consisting of four planes of atoms. Each of these
planes contains one type of atom, the ordering of the planes being
Se-Ga-Ga-Se. The symmetry of the crystal is hexagonal, with
the hexagonal axis perpendicular to the layers, and the atoms in
each plane are arranged in a regular close-packed hexagonal lattice.
The interatomic bonding is strong within each layer, but weak between
the layers, and this means that there are different ways of stacking the
layers which have almost the same energy. Four polytypes differing
in the stacking are experimentally known~\cite{kuhn75}, and the observed
form depends on the preparation method. There has been controversy
about which polytype is most stable.

Our aims in studying the crystal are to test our pseudopotentials
and to obtain an understanding of the electronic structure, so that
our main concern is with the strong intra-layer bonding. For this purpose,
it does not matter which polytype we examine, and we have chosen to
work with the $\beta$-phase, for which rather precise diffraction
data are available \cite{ben88}.
Because of the relation between adjacent layers
in this polytype, the primitive cell contains a total of eight atoms. The
crystal structure is characterized by the two lattice parameters
$a$ and $c$ and the two internal parameters $u$ and $v$. The $u$ and
$v$ parameters specify the positions of the Ga and Se planes respectively
along the $c$-axis. (For a detailed definition, see Wyckoff~\cite{wyc};
the $z$-coordinates for Ga and Se used in the paper of Benazeth
{\em et al.} \cite{ben88} correspond exactly to these $u$ and $v$ parameters.)

The calculations on $\beta$-GaSe were performed using a plane-wave cutoff of
250~eV. This cutoff was chosen on the basis of systematic tests
on the convergence of the total energy as a function of cutoff, which
showed that the residual error is roughly 1~meV per 8-atom cell.
Brillouin-zone sampling was performed using 18 Chadi-Cohen
points \cite{chc} in
the full zone. Again, tests done with different $k$-point sets indicate
that the residual error is no more than a few meV per cell.

We have performed a full structural relaxation of the crystal, which
gives the following equilibrium parameters (experimental values in
parentheses): $a$ = 3.64~\AA\ (3.750~\AA ), $c$ = 15.76~\AA\ (15.995~\AA ),
$u$ = 0.174 (0.1736) and $v$ = 0.602 (0.6015). The reasonably
close agreement with experiment fully confirms the adequacy of
our pseudopotentials. A slight underestimation of lattice parameters
is commonly found in DFT calculations, and we note that the calculated
lattice parameter in our earlier work on crystalline Ga was also too
low by $\sim$~3~\%. During the relaxation process, we have noted
the expected strong anisotropy in the energetics of the
crystal~\cite{abu95}, with changes of the $c$-parameter causing
much smaller energy changes than those of the $a$-parameter.

The electronic DOS calculated for the equilibrium structure is shown in
Fig. \ref{ds_s} (upper panel).
The calculations indicate that the material is a semiconductor,
with a band gap of 1.35\,eV (experimentally  about 2\,eV
\cite{mccanny77}). The electronic structure
of $\beta$-GaSe was studied many years ago by Schl\"{u}ter \cite{sch73}
using an
empirical model pseudopotential, and we find that his
band structure is in semi-quantitative agreement with our results.
So far as we know, ours are the first
{\em ab initio} calculations on the electronic structure of GaSe.

The valence part of the DOS consists of an isolated peak at --~14~eV
(bands 1-4),
a double peak at --~8~eV (bands 5-6 and 7-8),
and a broader distribution extending from
--~6~eV up to the valence band maximum (VBM) (bands 9-18).
A clearer understanding
of these features can be obtained by studying the partial electron
densities associated with chosen energy ranges. These densities
(Fig. \ref{ch_gases})
show that the peak at --~14~eV arises from Se(4$s$) states,
while the double peak arises from bonding and anti-bonding
states associated with Ga-Ga pairs, but with some weight on neighboring
Se atoms. These states can be regarded as made of Ga(4$s$) states, with a
strong admixture of other states. The broad peak in the DOS between
--~6~eV and the VBM appears to be mainly responsible for bonding in the
material, and consists of Ga(4$p$) and Se(4$p$) states. This
general analysis is consistent with what was found by Schl\"{u}ter
\cite{sch73}.
It is interesting to note that there have been recent first-principles
calculations on the closely related material InSe \cite{gom93}, and the
DOS calculated for that system shows all the same qualitative features
as ours.

We have also performed calculations on the Ga$_2$Se$_3$ crystal,
though here the experimental situation is rather unsatisfactory.
Again, there are a number of closely related structures, all of
which appear to be based on a defective zinc-blende structure, in
which one third of the cation sites are vacant. One of these
($\gamma$-Ga$_2$Se$_3$, which is
the high temperature phase) has full cubic symmetry, and the distribution
of vacancies appears to be random \cite{mikkelsen81}.

Our Ga$_2$Se$_3$ calculations were performed on a periodic system
having an 80-atom cell (Ga$_{32}$Se$_{48}$) consisting of twelve adjacent
zinc-blende cubes, with sixteen, randomly chosen,
vacant cation sites per cell.
For this system we have calculated the electronic density of states
(without relaxing the system) and the valence-charge density.
The calculations have been performed with
the same plane-wave cutoff as for $\beta$-GaSe.

The electronic DOS is shown in Fig. \ref{ds_s} (lower panel). Again, we
are dealing with a semiconductor, the gap in this case being about 0.6~eV.
The
valence DOS consists of the Se(4s) peak at about -14\,eV and a
broader distribution from -10 eV up to the energy gap.
The peaks in the DOS are not so well visible as in the case of
$\beta$-GaSe because of the randomness of $\gamma$-Ga$_2$Se$_3$.

\section{Liquid alloys}
\subsection{Structure and dynamics}
We have performed AIMD simulations of the liquid Ga-Se system for
the three compositions Ga$_2$Se, GaSe and Ga$_2$Se$_3$ at the
temperature 1300~K. All the simulations were done on a
repeating system of 60 atoms using $\Gamma$-point sampling,
with a plane-wave cut-off of 150~eV.
This cut-off is less than the one used for calculations
on the solids, but is adequate for simulations of the liquids. The
time step was taken to be 3~fs, and the Fermi smearing
was 0.2~eV. The density of these liquid mixtures is not
experimentally known, and the densities we use are obtained
from experimental values for $\ell$-Ga and $\ell$-Se. Data for
the density of $\ell$-Ga are available up to $\sim$~1000~K, and
the value at 1300~K can be estimated by a small extrapolation.
The case of $\ell$-Se is more tricky, since the boiling point under
atmospheric pressure is only 958~K. We have estimated the (hypothetical)
density at 1300~K by linear extrapolation of experimental
data between 490 and 958~K. Finally, the densities of
the liquid mixtures at 1300~K are obtained by linear interpolation
between the estimated values for $\ell$-Ga and $\ell$-Se at this
temperature. Some time after the work was started, a cross-check
became possible against recent neutron diffraction data on
$\ell$-Ga-Se at the same three compositions~\cite{gaseexp},
and this showed that our
estimated densities were correct to within better than 5~\% in
all cases. It is worth mentioning that linear interpolation
between {\it c}-Ga and {\it c}-Se predicts the densities of solid GaSe
and Ga$_2$Se$_3$ very accurately.

We initiated the simulations by starting from a typical
configuration of atoms taken from our previous $\ell$-Ga
simulations~\cite{ga95}, and replacing some of the Ga atoms by Se atoms.
At each composition, our simulated system was equilibrated
for 1~ps, and then a production run of 4~ps was generated.
The temperature and liquid structure were monitored throughout the
simulations, and the stability of these quantities indicated that
the 1~ps equilibration period was adequate.

The structure of the simulated liquid can be compared directly with
that of the real system through the static structure
factors $S_{\alpha \beta} (k)$.
These quantities, which give a measure
of the intensity of density fluctuations as a function of
wavevector $k$, are defined by:
\begin{equation}
\label{esf} S_{\alpha\beta}(k)=\bigl<\hat \rho_\alpha(\bbox{k})
\hat \rho^*_\beta (\bbox{k})\ \bigr>\ .
\end{equation}
Here the dynamical variable $\hat \rho_\alpha ({\bbox{k}})$
representing
the Fourier component of the atomic density of atoms type $\alpha$
at wavevector $\bbox{k}$ is given by:
\begin{equation}
\label{sf}
    \hat \rho(\bbox{k})=N_\alpha^{-1 / 2}\,\sum_{i=1}^{N_\alpha} \,
    \exp(i\bbox{k \cdot r}_i)\ ,
\end{equation}
where $\bbox{r}_i$ is the position of atom $i$ and $N_\alpha$ is the
number of
atoms of type $\alpha$ in the system.  The angular brackets in Eq.
(\ref{esf})
denote the thermal average, which in practice is evaluated as the
time average over the duration of the simulation.
The structure factors were calculated
for vectors {\bf k} compatible with the periodic
boundary conditions.
In practical
calculations of $S_{\alpha\beta}(k)$, we also average over $\bbox{k}$
vectors
having the same magnitude.
The neutron weighted structure factor $S_n (k)$ measured in a neutron
diffraction experiment is given by:
\begin{equation}
S_n(k)=\frac{\sum_{\alpha\beta}\sqrt{c_{\alpha}c_{\beta}}{b_{\alpha}b_
{\beta}}S_{\alpha \beta}(k)}{\sum_{\alpha}c_{\alpha}b_{\alpha}^{2}} \ ,
\end{equation}
where $c_{\alpha}$ and $b_{\alpha}$ are the concentration and the
neutron
scattering length of species $\alpha$ respectively. In the
present work, we took the scattering lengths for Ga and Se to be
$b_{\rm Ga}$=7.288~fm and $b_{\rm Se}$=7.97~fm~(Ref.~\onlinecite{gaseexp}).
We shall also need to refer to the total structure factor $S (k)$,
which is defined by:
\begin{equation}
S(k)=\sum_{\alpha\beta}\sqrt{c_{\alpha}c_{\beta}} S_{\alpha \beta}(k) \ .
\end{equation}
In fact, $S(k)$ is virtually identical to $S_n (k)$ for the present
systems, since $b_{\rm Ga}$ and $b_{\rm Se}$ are almost the same.

The neutron-weighted liquid structure factors $S_n (k)$ calculated for the
three compositions are compared with the recent neutron
diffraction data~\cite{gaseexp} in Fig. \ref{sfn}. The agreement is
good for wavevectors above $\sim$~3~\AA$^{-1}$, and is reasonably
satisfactory  below that. Since the main discrepancies are at
low $k$, it is possible that the rather small size of our
simulated system may be influencing the results; limitations
of computer power prevent us from going to larger systems
at present. The main changes in the experimental $S_n (k)$
with increasing Se content are the appearance of a small peak at
$\sim$~1~\AA$^{-1}$, a shift and enhancement of the peak at
$\sim$~2~\AA$^{-1}$ and an increase in amplitude and shift in phase
of the oscillations beyond $\sim$~3~\AA$^{-1}$. All these
effects are correctly reproduced in the simulations.

The origin of the features in $S_n (k)$ can be understood by
examining the partial structure factors $S_{\alpha \beta} (k)$
shown in Fig. \ref{sffz}.
These results make it clear that the main peak in $S_n(k)$
at $\sim$~3~\AA$^{-1}$
arises from a superposition of almost coincident peaks
in all three $S_{\alpha \beta} (k)$. The peak at $\sim$~2~\AA$^{-1}$
comes mainly from a peak in $S_{\rm Se-Se} (k)$, which is
partially cancelled by a trough in $S_{\rm Ga-Se} (k)$;
both of these features shift to lower $k$ as the Se content
is increased. The appearance of the small peak at $\sim$~1~\AA$^{-1}$
is due mainly to a growth of Ga-Se correlations with increasing
Se content.

The real-space structure of the liquid can be understood through the
partial radial distribution functions (RDFs) $g_{\alpha \beta} (r)$
(Fig.~\ref{rdf}), and the coordination numbers of their
first peaks (table~\ref{table1}). One can notice very strong
similarities between the structures of solids and the corresponding
liquids. Not only are the interatomic  distances very similar but also
the coordination numbers are very close.

It is useful to relate the forms of the RDFs to the corresponding
crystal structures.
The main peak in $g_{\rm Ga-Se}$ arises from
directly bonded Ga-Se pairs, and its distance is close to
the corresponding bond length in the crystals. The main peak in
$g_{\rm Se-Se}$ comes from pairs of Se atoms bonded to the
same Ga atom, and its distance remains unchanged for the same
reason that the Se-Se distance is almost the same in the GaSe and
Ga$_2$Se$_3$ crystals. We note, however, the formation of a
small peak in $g_{\rm Se-Se}$ at $\sim$~2.4~\AA\
at the Ga$_2$Se$_3$ composition, which is due to
Se atoms directly bonded to each other; this is confirmed by
the fact that 2.4~\AA\ is almost exactly the bond length in
crystalline \cite{wyc}
and liquid Se~\cite{inui90}. At low Se content, $g_{\rm Ga-Ga}$
consists mainly of a peak at $\sim$~2.6~\AA, whose height decreases
strongly on going to Ga$_2$Se$_3$, with the growth of a second
peak at $\sim$~3.8~\AA. This is related  to the presence of
Ga-Ga bonds in crystalline GaSe and their absence in Ga$_2$Se$_3$.

We have studied the diffusion behavior of Ga and Se by calculating
the time-dependent mean square displacements for the two species.
These show the usual liquid-like behavior, with a rapid transition
to a linear dependence on time after about 0.1~ps. The
self-diffusion coefficients $D_{\rm Ga}$ and $D_{\rm Se}$ are obtained
in the usual way from the slope, and the calculated values are
shown in Table \ref{table2}. The results show that both diffusion
coefficients decrease rather strongly with increasing Se content,
presumably because of the formation of a fairly stable covalently
bonded network. Interestingly, though, recent AIMD simulations
that we have made on pure $\ell$-Se~\cite{aimdse} show that its diffusion
coefficient at 1375~K has a high value of
1.5$\times 10^{-4}$~cm$^2$s$^{-1}$. We
conjecture that $D_{\rm Se}$ has a minimum at the composition
Ga$_2$Se$_3$ and increases rapidly thereafter.

\subsection{Electronic properties}
The total electronic densities of states (DOS) of the GaSe and
Ga$_2$Se$_3$ liquids are compared with our results for the
corresponding crystals in Fig.~\ref{ds}. The comparison shows the very
close resemblance of the solid and liquid phases, which might be
expected from the similarity of their short-range order. In both
cases, the liquid-state DOS is simply a rather broadened version of the
crystal DOS but shifted to higher energies.
The main difference for GaSe is the replacement
of the band gap by a minimum in the DOS. The variation of
electronic structure with composition is illustrated in Fig.~\ref{dsall},
where we include the DOS for $\ell$-Ga at 982~K taken from our
previous work~\cite{ga95}. The $\ell$-Ga system is highly metallic, and has a
free-electron-like DOS. The change on going to Ga$_2$Se is very
marked, with the appearance of the feature due to Se(4$s$) states and
the formation of the broad two-component distribution consisting
of hybridized Ga(4$s$/$p$)-Se(4$p$) states characteristic also of
GaSe and Ga$_2$Se$_3$. The DOS at the Fermi level shows a
monotonic decrease with increasing Se content, reaching a zero value
at the Ga$_2$Se$_3$ composition.

The close resemblance between the electronic structure of the
solid and liquid is confirmed by a study of the electron density
distribution. A useful way to do this in the liquid is to examine
the density on a plane passing through two neighboring Ga atoms
and a Se atom neighboring one of these Ga atoms. We show in
Fig.~\ref{ch_gasel}
the energy resolved partial densities for a typical configuration
in $\ell$-GaSe, which can be directly compared with the
corresponding results for the crystal (Fig.~\ref{ch_gases}). All
the characteristic features found in the solid -- the Se(4$s$)
states (bands 1-30), the bonding and anti-bonding states on the Ga pair
(bands 31-45 and 46-60), and the
Ga-Se and Ga-Ga bonding states associated with the upper part
of the valence band (bands 61-135) -- are clearly visible in
the liquid. We find
a similar close resemblance for solid and liquid Ga$_2$Se$_3$,
as illustrated in Fig.~\ref{ch_ga2se3}.

The relation between the electron density distribution in the solid
and liquid can be pursued to a more quantitative level. To do this,
we have chosen particular Ga-Ga and Ga-Se pairs in liquids of
different compositions and in the solids, and plotted the
electronic density along the bonds. Since the distances between neighbors
vary in the liquid, we reduce lengths to a common scale by
dividing by the interatomic distance. Some typical
comparisons are shown in Fig.~\ref{bonds}. The agreement between the
solid-state and liquid-state curves, even for different
compositions, is remarkable, and indicates that there is almost
no change in the short-range electronic structure between the
different phases.

We have calculated the electronic d.c. conductivity of the three
liquid phases using the Kubo-Greenwood approximation~\cite{mott}.
The technique used is the one described in several previous
papers~\cite{stich89,ga95}.
The calculations were done by averaging over the full duration of
each simulation, and used $\Gamma$-point sampling, which is expected
to be adequate for present purposes. Fermi-Dirac thermal occupation
numbers were included in the calculations. The results are
compared with the very recent experimental measurements of
Lague and Barnes~\cite{gaseexp} in Fig. \ref{cond}. In this graph we
also include results for pure Ga \cite{ga95}.
Given the approximations involved,
the agreement is as good as can be expected. The dramatic
decrease of conductivity is, of course, associated with the
reduction of DOS at the Fermi level already described.

\section{Discussion and conclusions}
The good agreement between the calculated equilibrium structure
of the GaSe crystal with available data, as well
as the satisfactory comparisons of the liquid structure factors
with recent diffraction data, give confidence in the realism
of our AIMD simulations. One of the main findings from these
simulations is the very close relation between the
properties of the solids and the liquids. The positions of the peaks
in the liquid RDFs for GaSe and Ga$_2$Se$_3$ are close to the
interatomic distances in the crystals, and the coordination
numbers are also similar. As suggested in the Introduction,
the bonding in the liquids appears to be a mixture of metallic
and covalent, with the structure dominated by Ga-Ga and Ga-Se bonding
in the composition range we have examined. These findings about the
similarity of solid and liquid and the nature of the bonding are
in line with indications from early diffraction work on the
related Ga-Te and In-Te liquids~\cite{takeda83}.

Clearly at some point in the composition range, direct bonding between
Se atoms must become important, since the structure of pure $\ell$-Se is
entirely determined by this bonding. According to our simulations,
this direct bonding is insignificant in the range up Ga$_2$Se$_3$.
However, there is clear evidence in the $g_{\rm Se - Se}$ RDF
for the beginning of this effect at the Ga$_2$Se$_3$ composition.
It is interesting here to compare with our recent AIMD simulations on
the $\ell$-Ag-Se system~\cite{agse95},
where we found that direct Se-Se bonding begins at the
Ag$_2$Se composition, and rapidly becomes a dominant effect for higher
Se contents. This strongly suggests that the point at which Se-Se
bonding begins is determined by the maximum valency of the other
component. In this context, {\em ab initio} simulations or
diffraction measurements on systems such as Cu-Se and Cd-Se would be
extremely interesting.

The strong similarity of solid and liquid is also very clear
from our results for the electronic DOS, where we have shown
that for both GaSe and Ga$_2$Se$_3$ the liquid DOS is essentially a
broadened version of the solid DOS, except for the disappearance
of the band gap in the case of GaSe. Interestingly, the band gap
does not disappear for $\ell$-Ga$_2$Se$_3$. This behavior of the
DOS is closely related to the composition dependence of the electrical
conductivity, which, according to our rather limited results, decreases
rapidly and
monotonically as one passes from $\ell$-Ga to $\ell$-Ga$_2$Se$_3$,
as is also found experimentally.

\section*{Acknowledgments}
The work of JMH is supported by EPSRC grant GR/H67935. The computations
were performed on the Fujitsu VPX~240 at Manchester Computer Centre
under EPSRC grant GR/J69974. Analysis of the results made use of
distributed hardware provided by EPSRC grant GR/36266. We gratefully
acknowledge useful discussions with J.E.~Enderby, A.C.~Barnes, S.B.~Lague
and F.~Kirchhoff.

\begin{figure}
\caption{Structure of the $\beta$-GaSe crystal. Dark and light
spheres represent Ga and Se respectively.}
\label{b-gase}
\end{figure}
\begin{figure}
\caption{Calculated electronic DOS for $\beta$-GaSe (upper panel)
and $\gamma$-Ga$_2$Se$_3$ (lower panel). For presentation
purposes, the calculated DOS are convoluted with a Gaussian of
width 0.1~eV. The horizontal scale representsthe difference of the
energy E and the Fermi energy E$_F$.}
\label{ds_s}
\end{figure}
\begin{figure}
\caption{The band-resolved and total charge densities for
the $\beta$-GaSe crystal (units: 10$^{-2}$~\AA$^{-3}$). Ga and Se sites
are marked by filled circles and squares respectively. Lengths are
marked in \AA\ units. The first four panels show bands resolved charge
densities corresponding to features in the DOS (for meaning of bands
see text).
The last panel shows the total charge density.}
\label{ch_gases}
\end{figure}
\begin{figure}
\caption{Simulated (solid line) and experimental (circles) neutron-weighted
structure factors of liquid Ga-Se at three compositions. The vertical
scale refers to the GaSe results, with Ga$_2$Se and Ga$_2$Se$_3$
results shifted respectively up and down by one unit.}
\label{sfn}
\end{figure}
\begin{figure}
\caption{Faber-Ziman partial structure factors $S_{\alpha \beta} (k)$
and total structure factor (see text) obtained from simulations
of liquid Ga-Se at three compositions.}
\label{sffz}
\end{figure}
\begin{figure}
\caption{Partial radial distribution functions $g_{\alpha \beta} (r)$
obtained from simulations of liquid Ga-Se at three
concentrations. Solid and dashed vertical lines indicate interatomic
distances in the $\beta$-GaSe and $\gamma$-Ga$_2$Se$_3$ crystals
respectively.}
\label{rdf}
\end{figure}
\begin{figure}
\caption{Comparison of the electronic DOS for
solid (~\protect\rule[1mm]{10mm}{0.2mm}~) and
liquid (~$- \, - \, - \, -$~) GaSe and Ga$_2$Se$_3$.}
\label{ds}
\end{figure}
\begin{figure}
\caption{Comparison of the electronic DOS calculated for $\ell$-Ga and
$\ell$-Ga-Se at three compositions.}
\label{dsall}
\end{figure}
\begin{figure}
\caption{Band-resolved and total electron densities for
$\ell$-GaSe (units: 10$^{-2}$~\AA$^{-3}$).
Ga and Se sites
are marked by filled circles and squares respectively. Lengths are
marked in \AA\ units. The first four panels show bands resolved charge
densities corresponding to features in the DOS (for meaning of bands
see text).
The last panel shows the total charge density.}
\label{ch_gasel}
\end{figure}
\begin{figure}
\caption{Total electron density for solid (left panel) and liquid
(right panel) Ga$_2$Se$_3$ (units: 10$^{-2}$~\AA$^{-3}$). Lengths
are marked in \AA\ units.}
\label{ch_ga2se3}
\end{figure}
\begin{figure}
\caption{Electron density (units: 10$^{-2}$~\AA$^{-3}$) along
lines joining neighboring Ga-Ga and Ga-Se pairs in various solid
and liquid phases.}
\label{bonds}
\end{figure}
\begin{figure}
\caption{Comparison of calculated (open circles) and experimental
(filled diamonds) electronic
conductivities for liquid Ga$_{1-x}$Se$_x$ alloys.}
\label{cond}
\end{figure}

\begin{table}
\caption{The average coordination numbers for liquid Ga-Se alloys and
positions of the first peaks in the radial distribution fuction,
in brackets the interatomic distances and the coordination numbers
of the corresponding solid phase. Distances in \AA.}
\label{table1}
\begin{tabular}{cccccccc}
liquid&$r_{\rm Ga-Ga}$&$n_{\rm Ga-Ga}$&$r_{\rm Ga-Se}$&
$n_{\rm Ga-Se}$ &$n_{\rm Se-Ga}$&
$r_{\rm Se-Se}$&$n_{\rm Se-Se}$ \\ \hline
Ga$_2$Se & 2.54 & 2.41 & 2.44 & 1.46 & 2.92 & 3.95$^a$ & 6.2$^a$  \\
GaSe & 2.51 (2.44) & 1.12 (1) & 2.44 (2.45)
& 2.66 (3) & 2.66 (3) & 3.95$^a$ (4.75$^a$) & 8.0$^a$ (9$^a$)  \\
Ga$_2$Se$_3$ & 2.46 (3.85$^a$) & 0.62 (0)& 2.43 (2.35)
& 3.39 (4) & 2.26 (2.67) & 3.95$^a$ (3.85$^a$) & 9.0$^a$ (12$^a$) \\
\end{tabular}
$^a$ these numbers correspond to the second coordination shell.
\end{table}
\begin{table}
\caption {Calculated diffusion coefficients (in $10^{-5}$cm$^2$s$^{-1}$)
for Ga-Se liquids.}
\label{table2}
\begin{tabular}{ccccc}
liquid&$D_{Ga}$&$D_{Se}$ \\ \hline
Ga$_2$Se & 9.8 & 5.8 \\
GaSe & 4.6 & 3.0 \\
Ga$_2$Se$_3$ & 2.2 & 2.5  \\
\end{tabular}
\end{table}

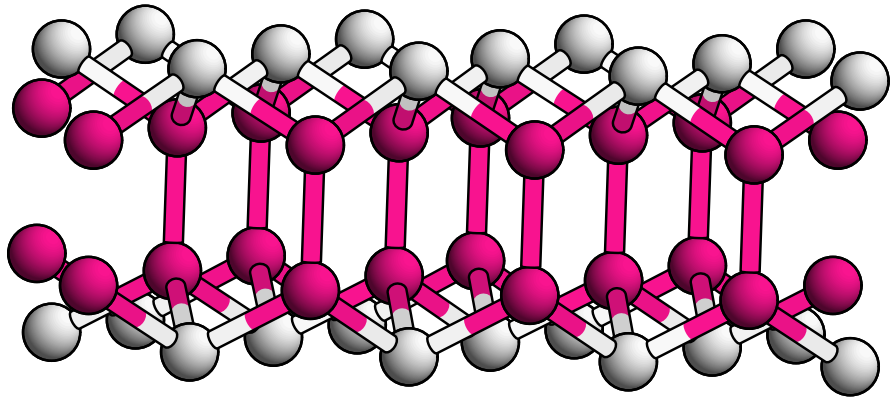
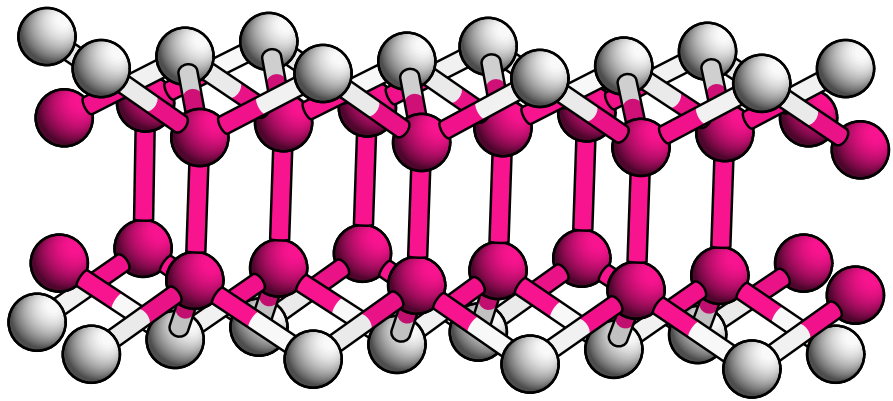
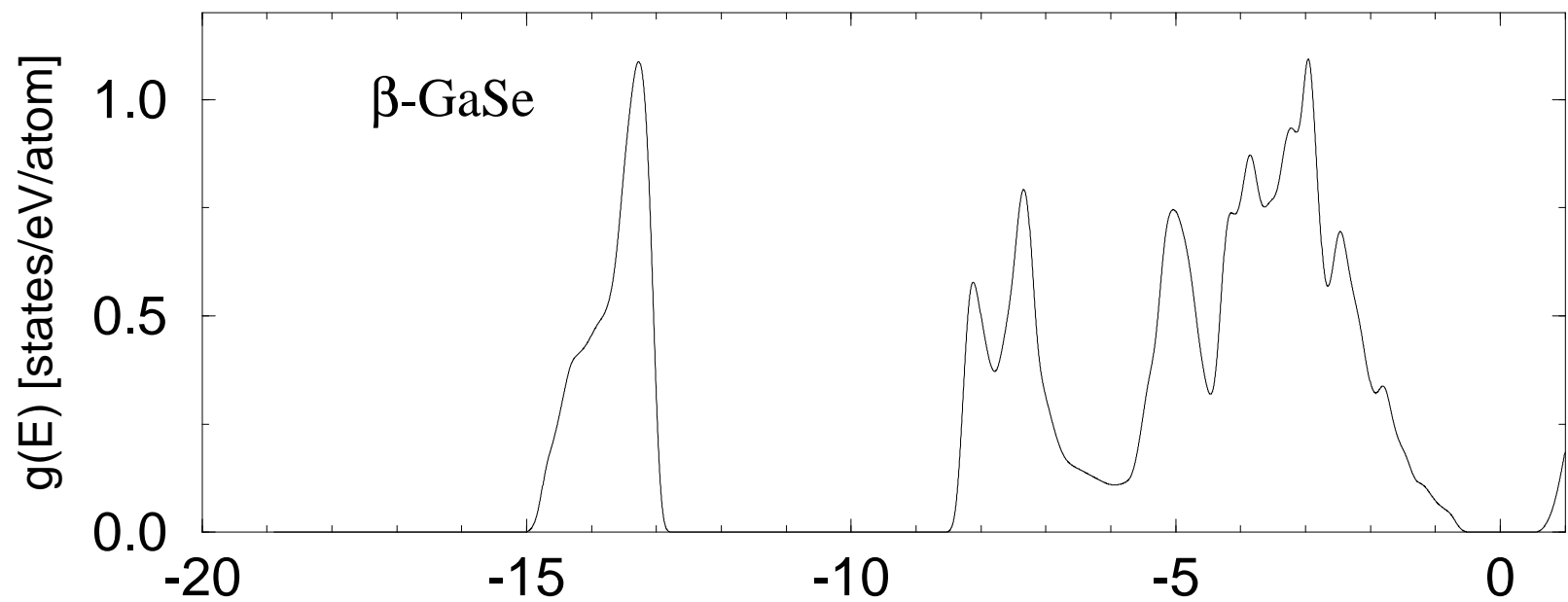
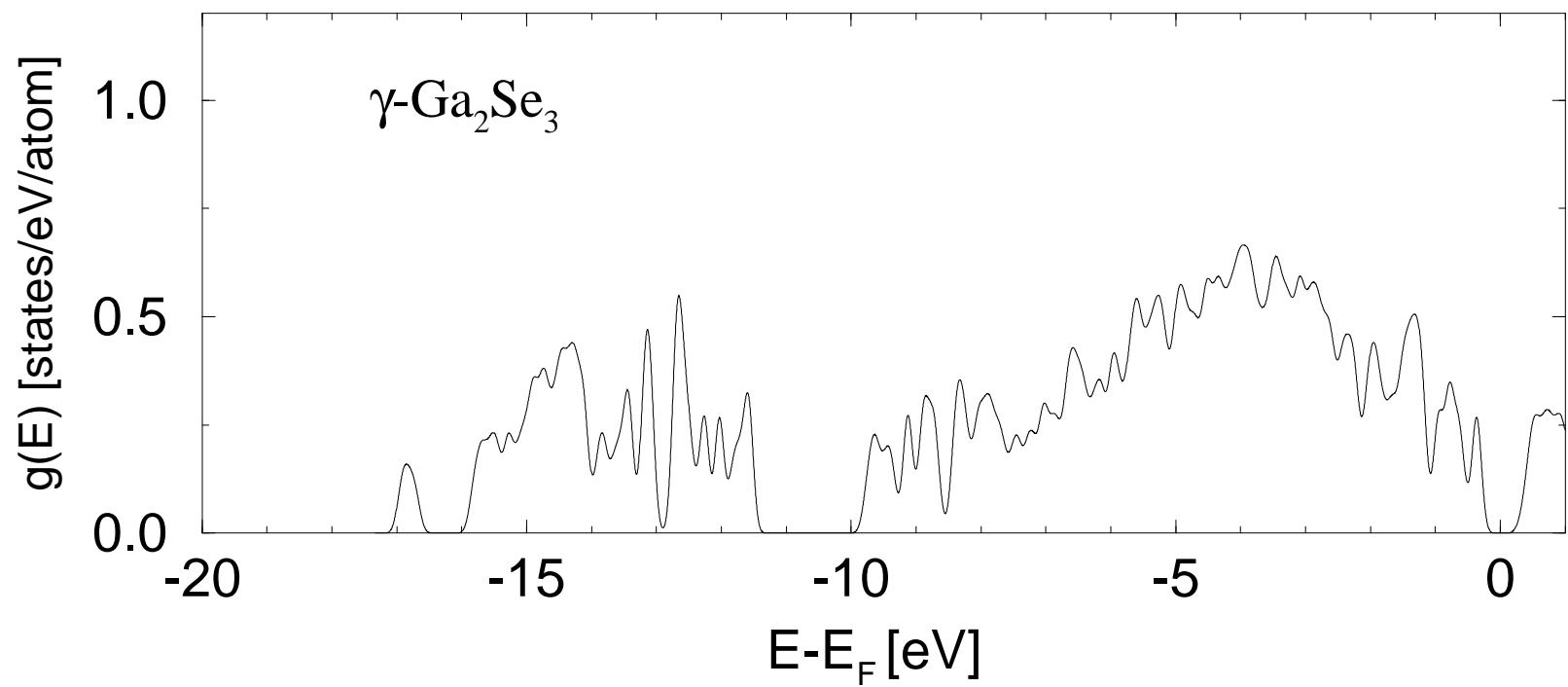
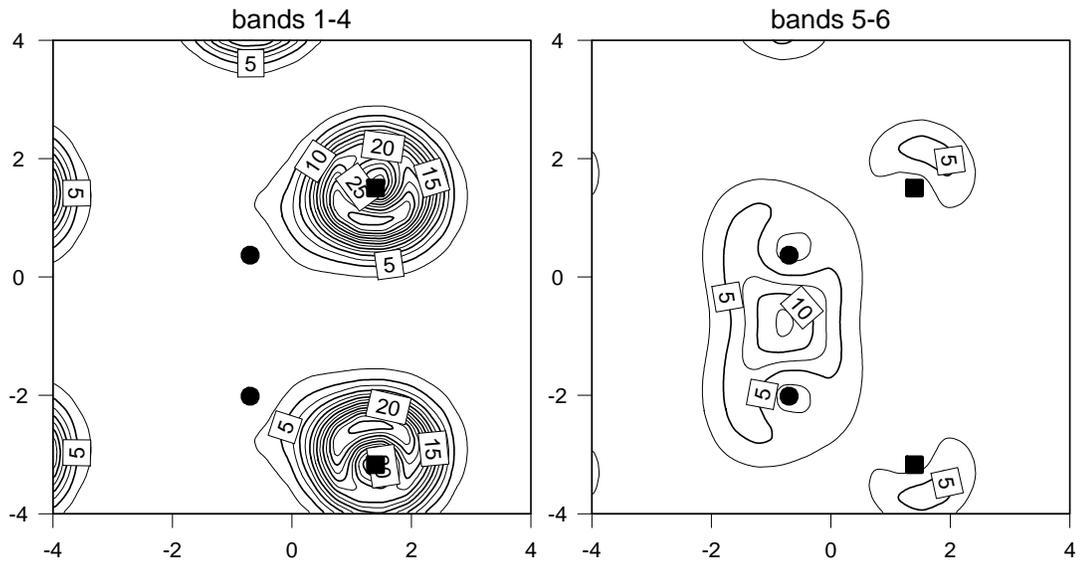
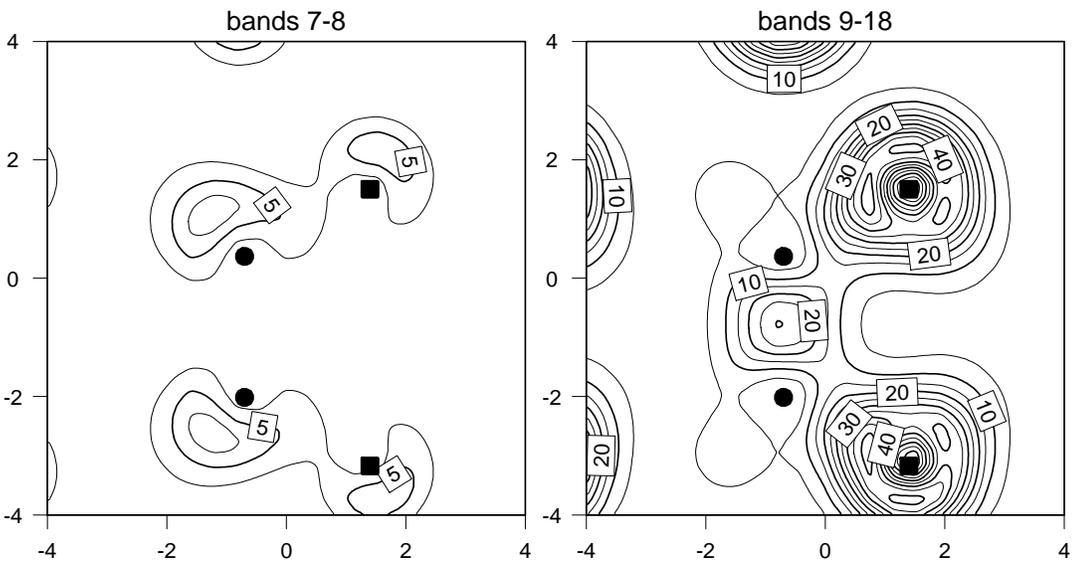
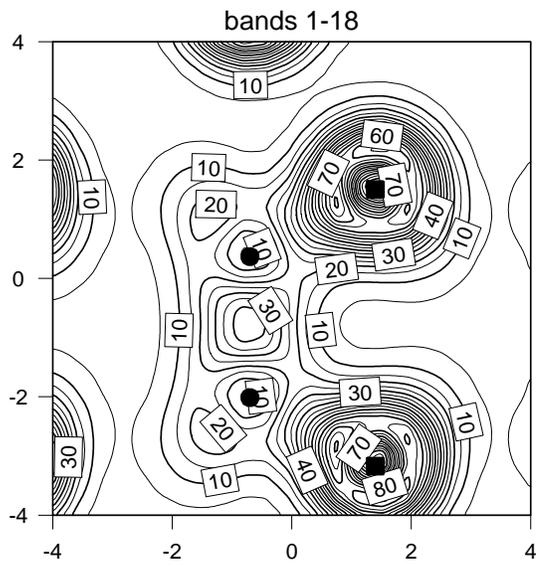
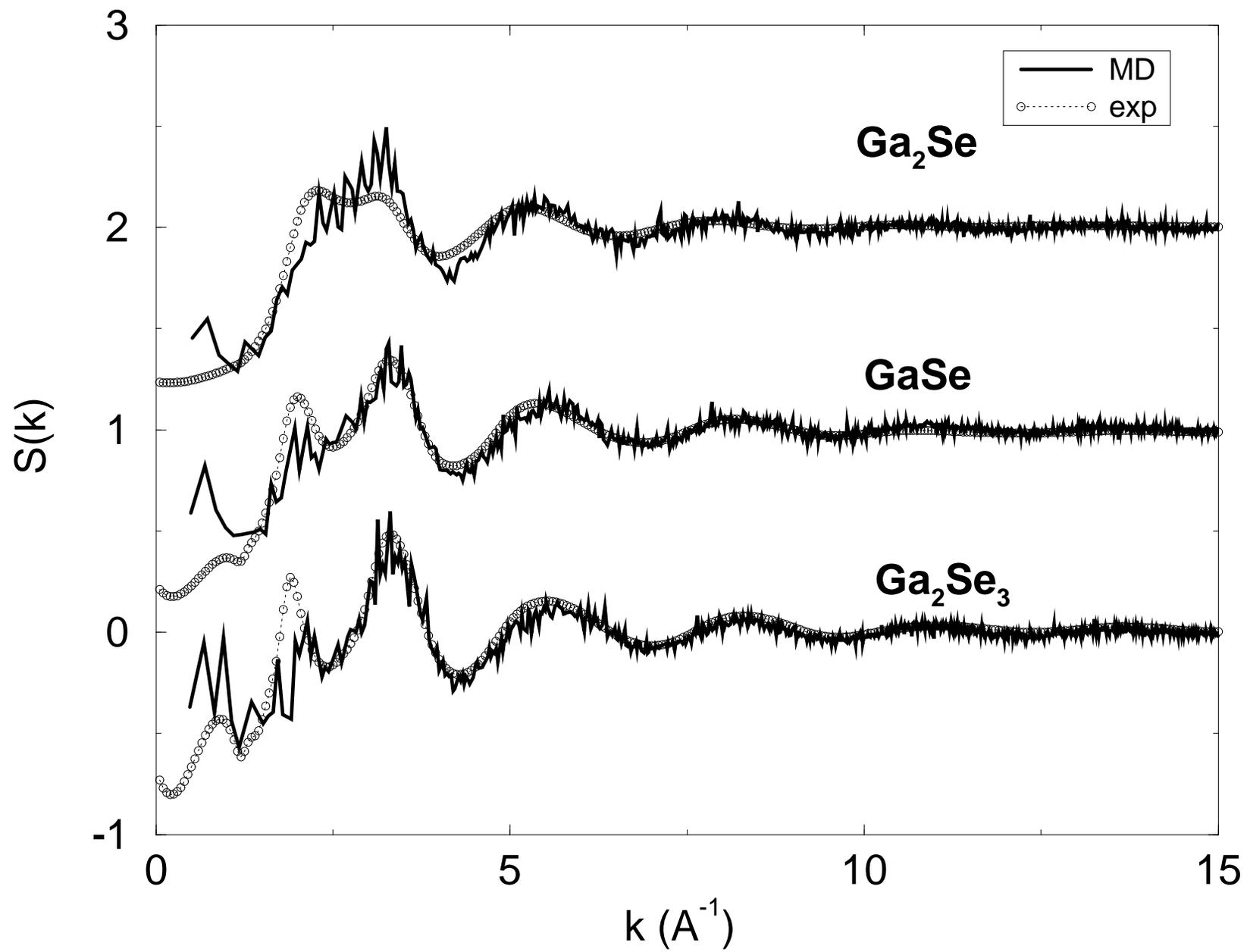
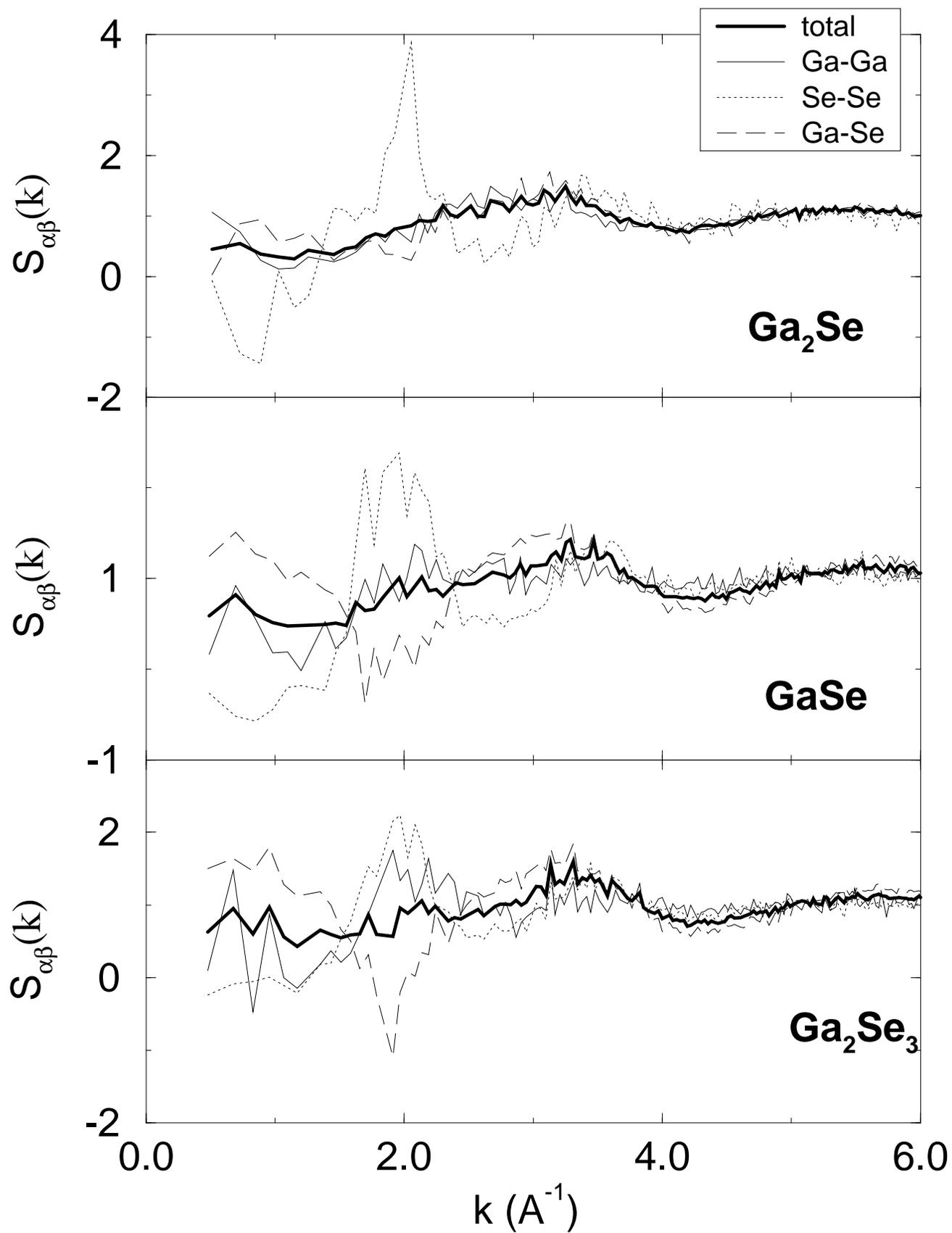
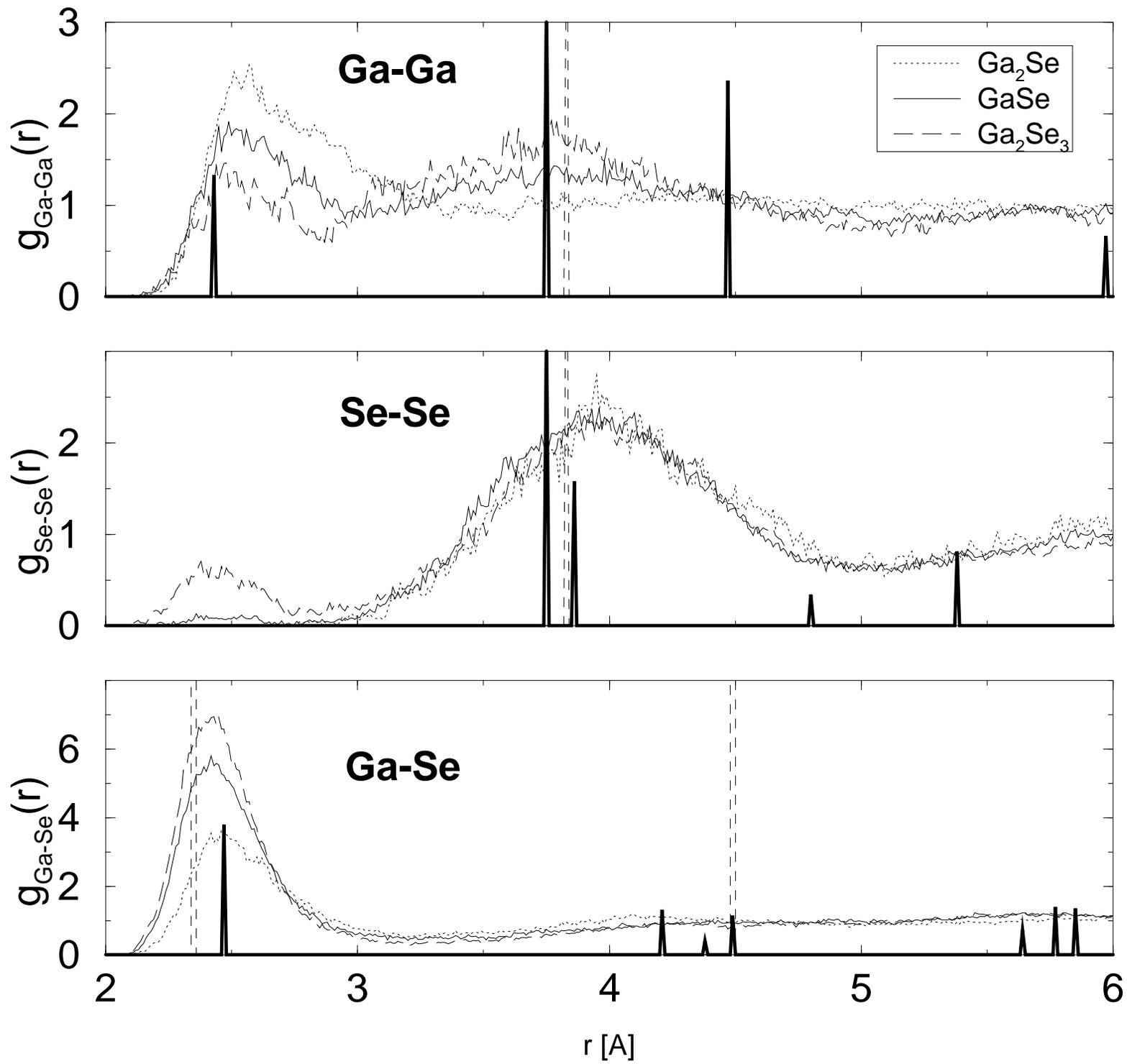
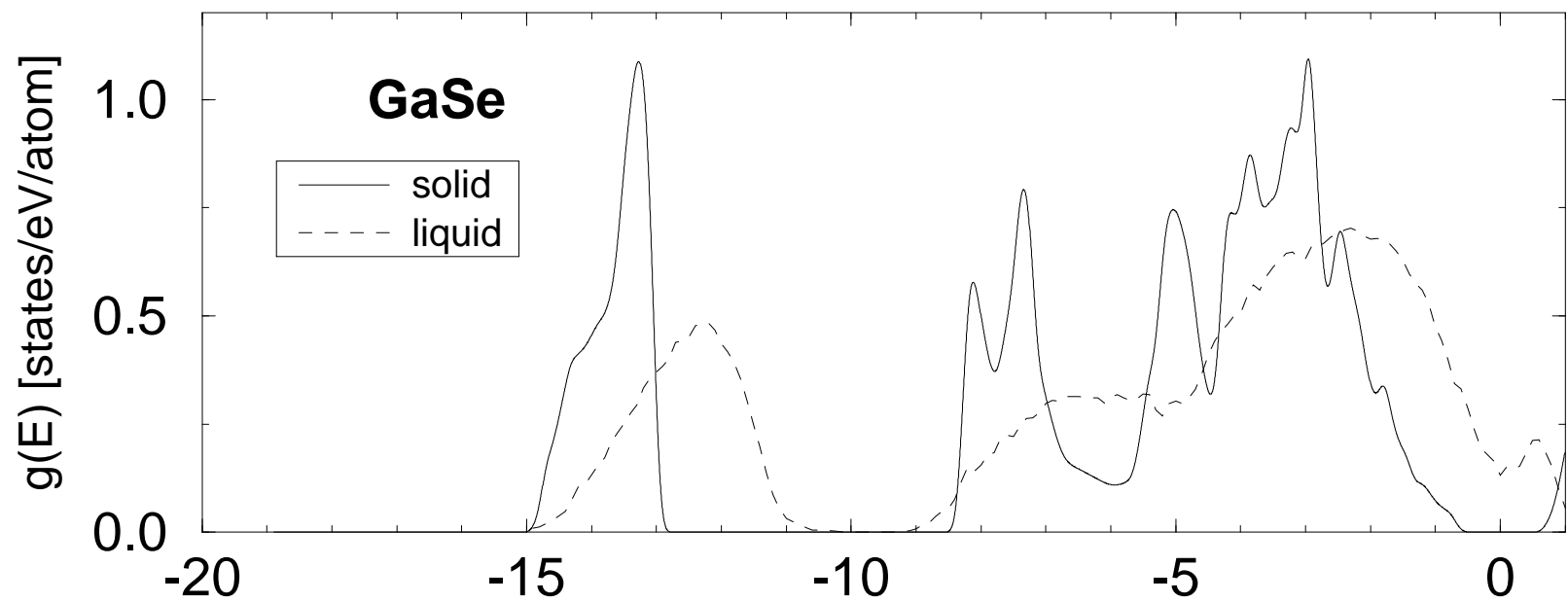
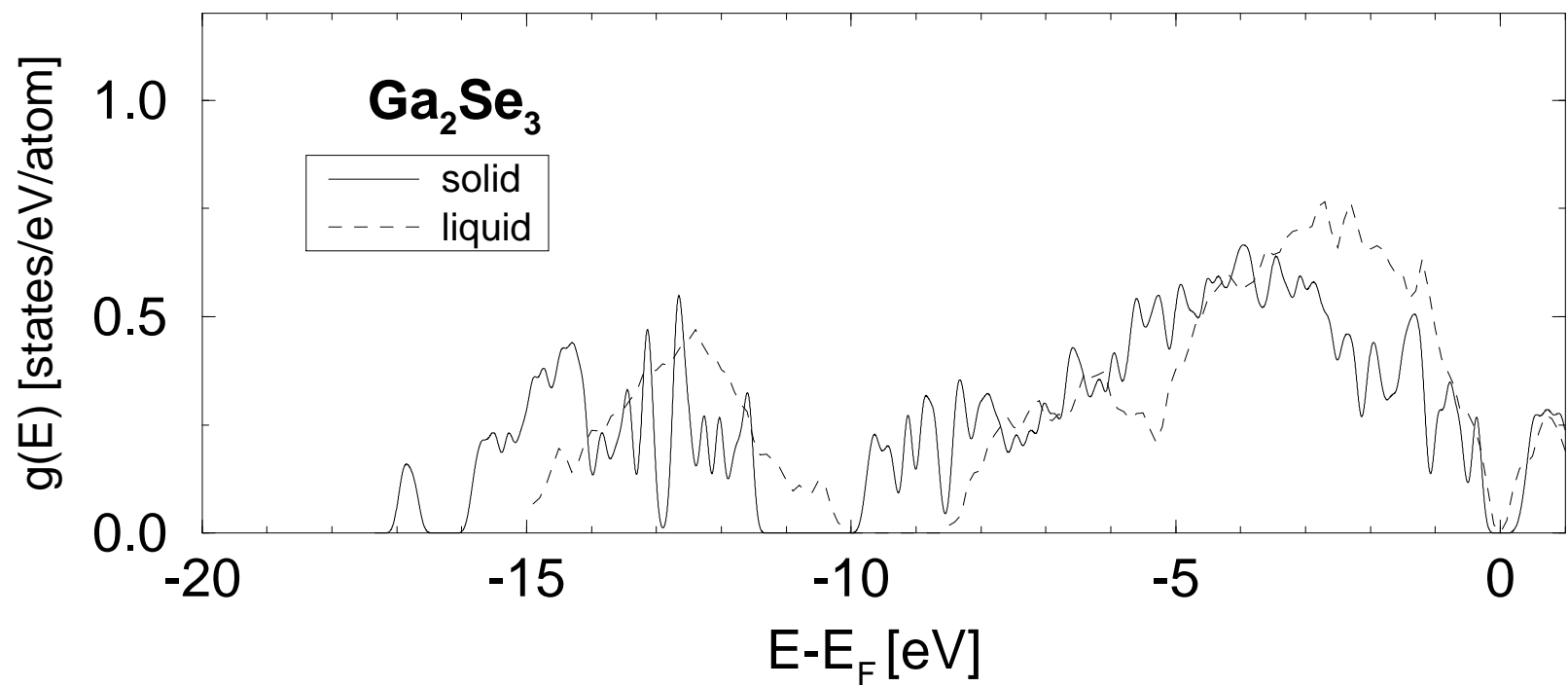
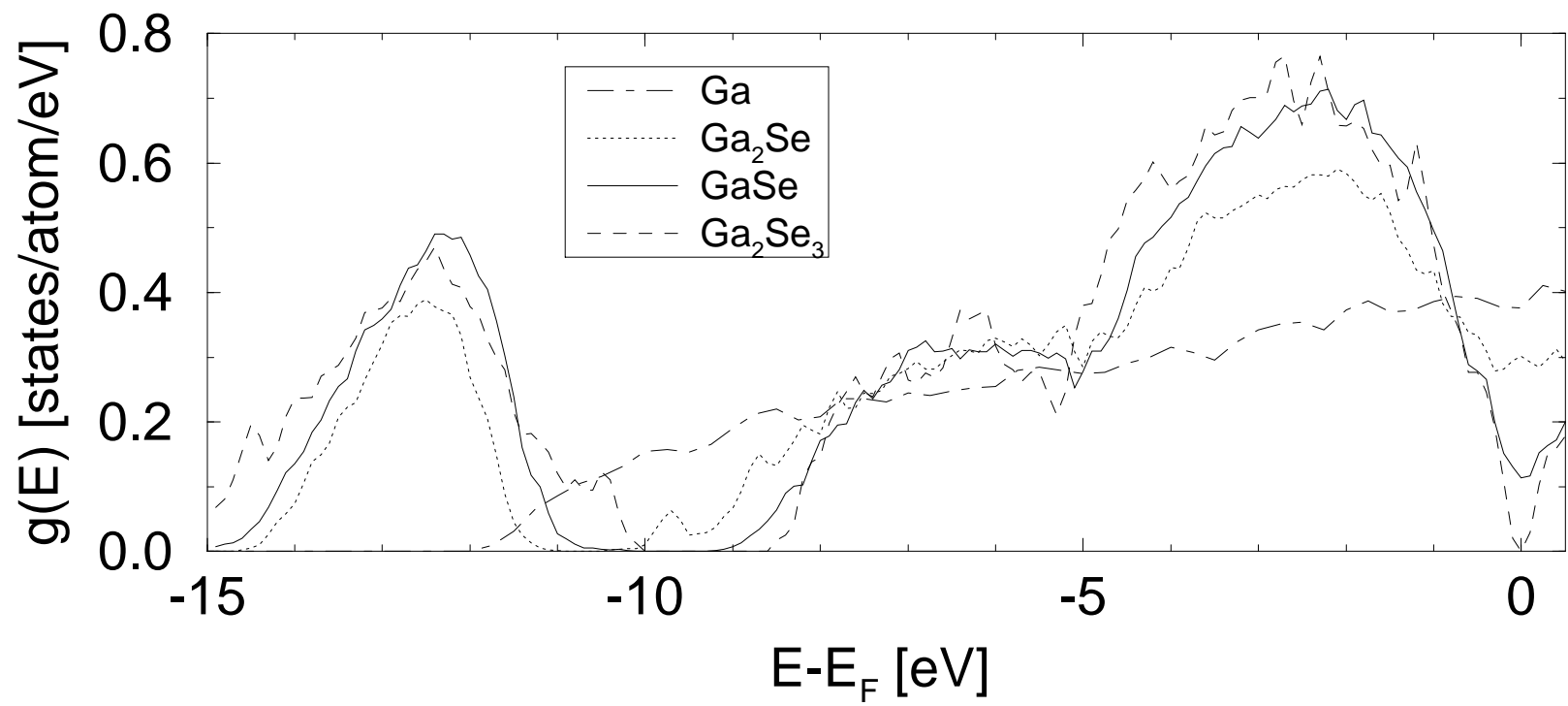
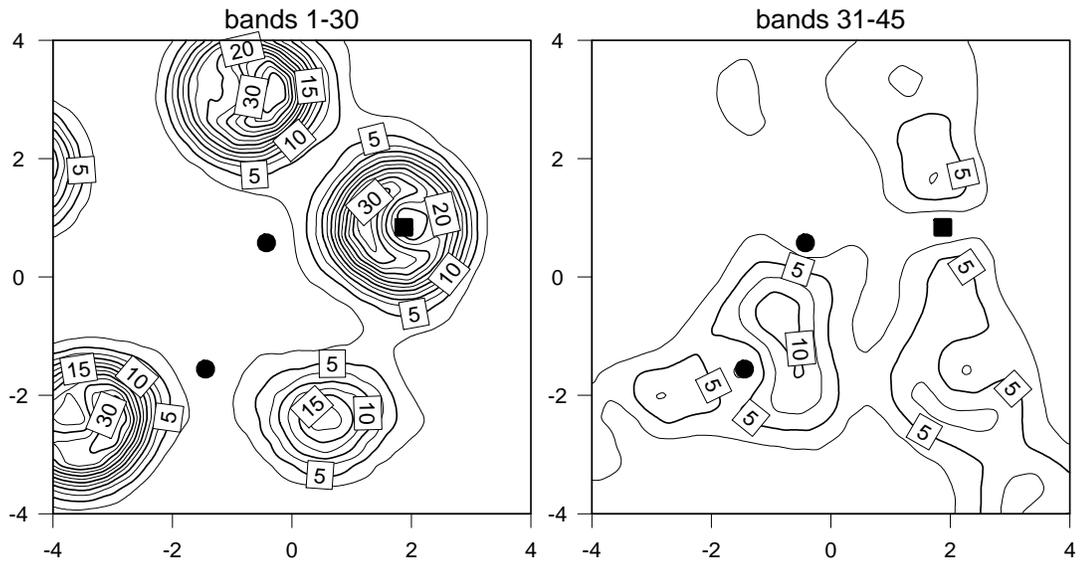
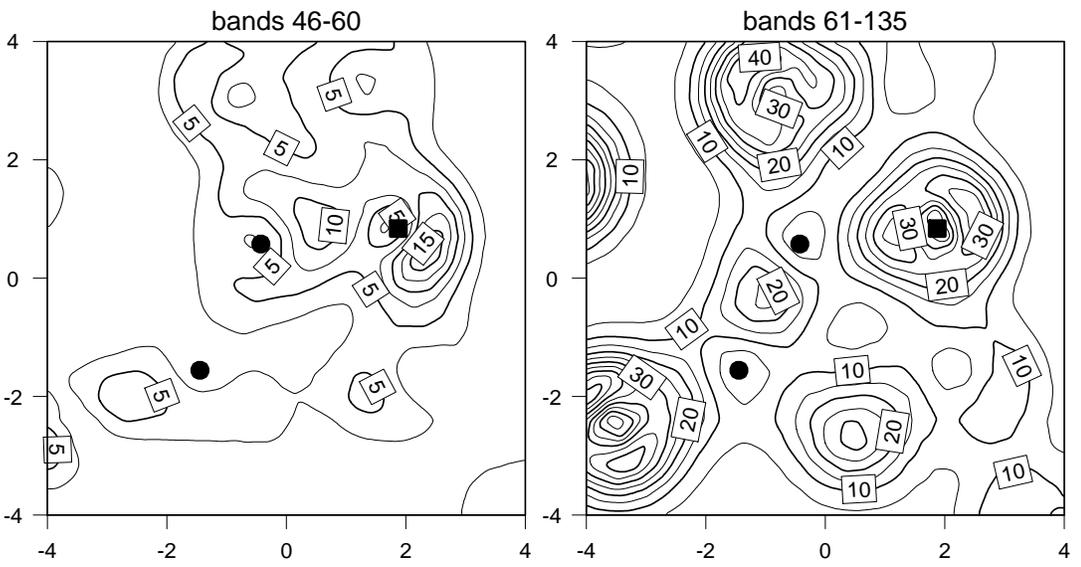
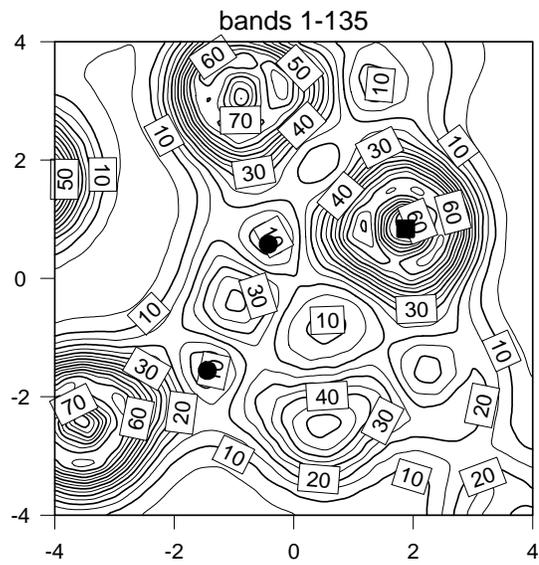
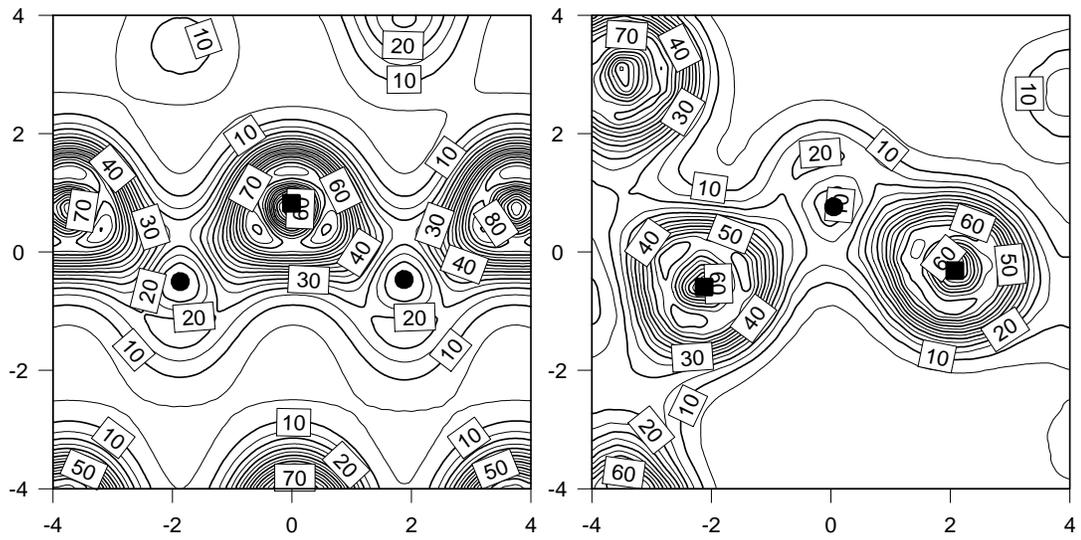
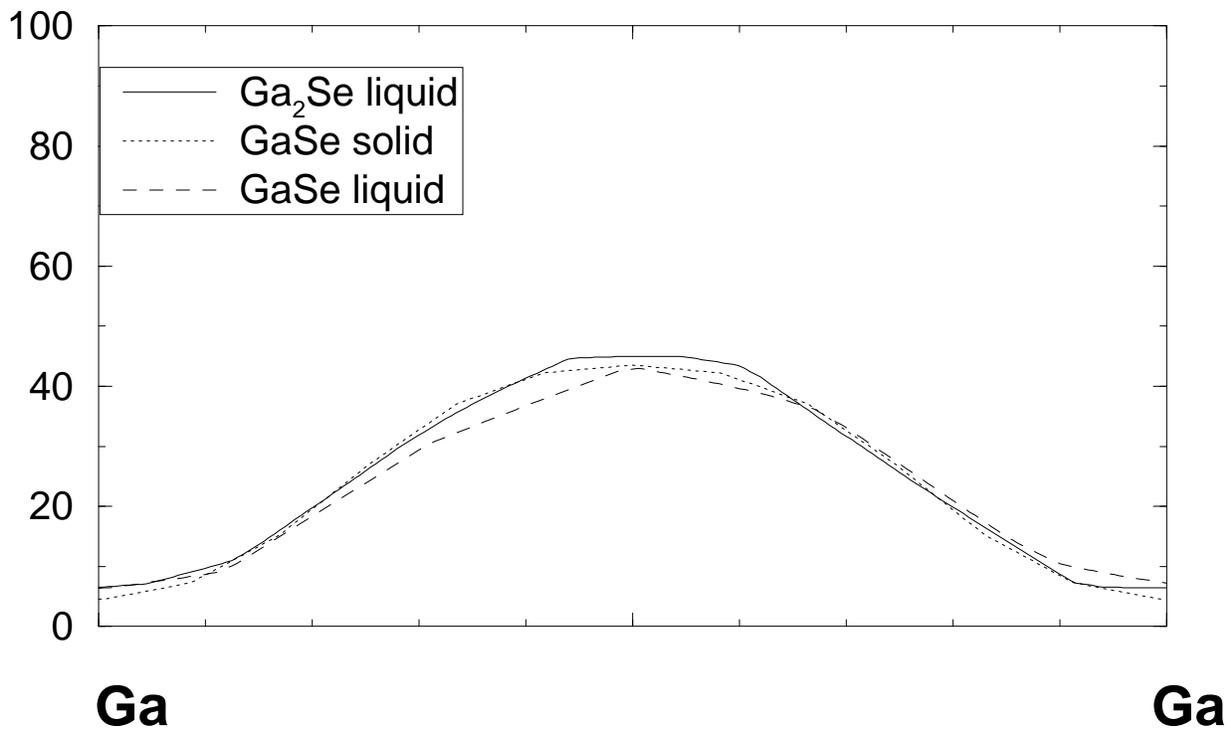
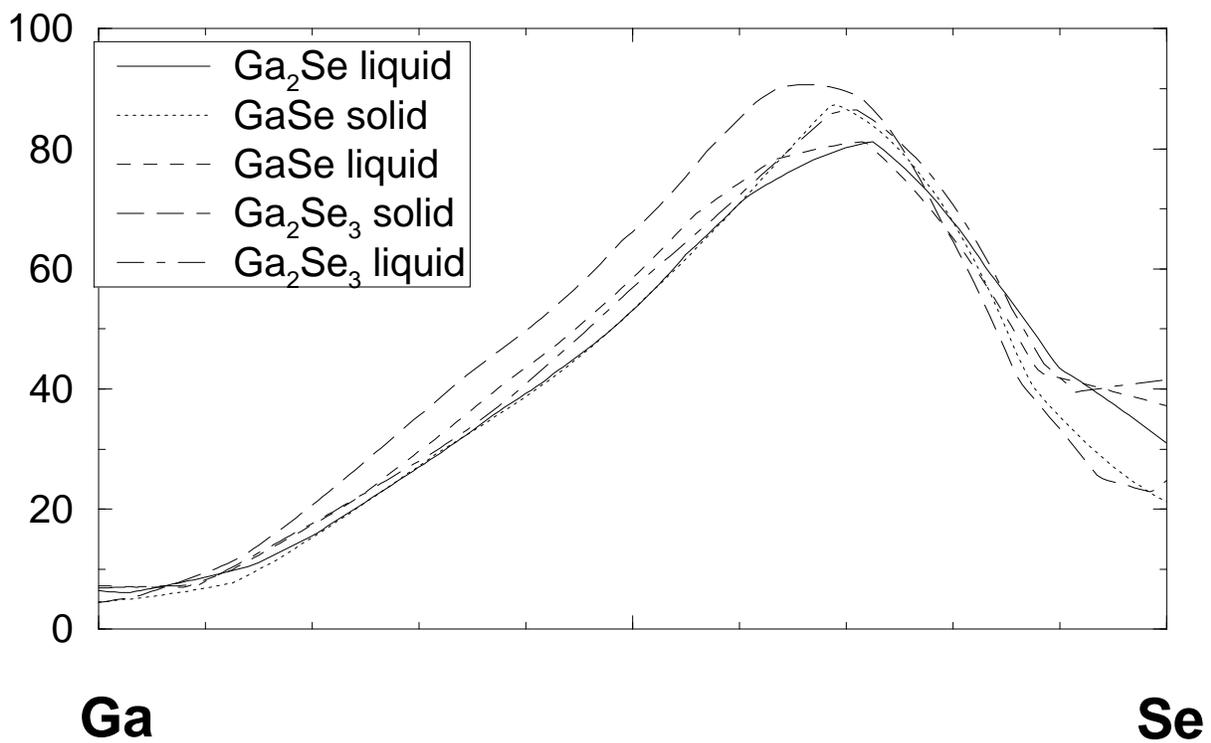
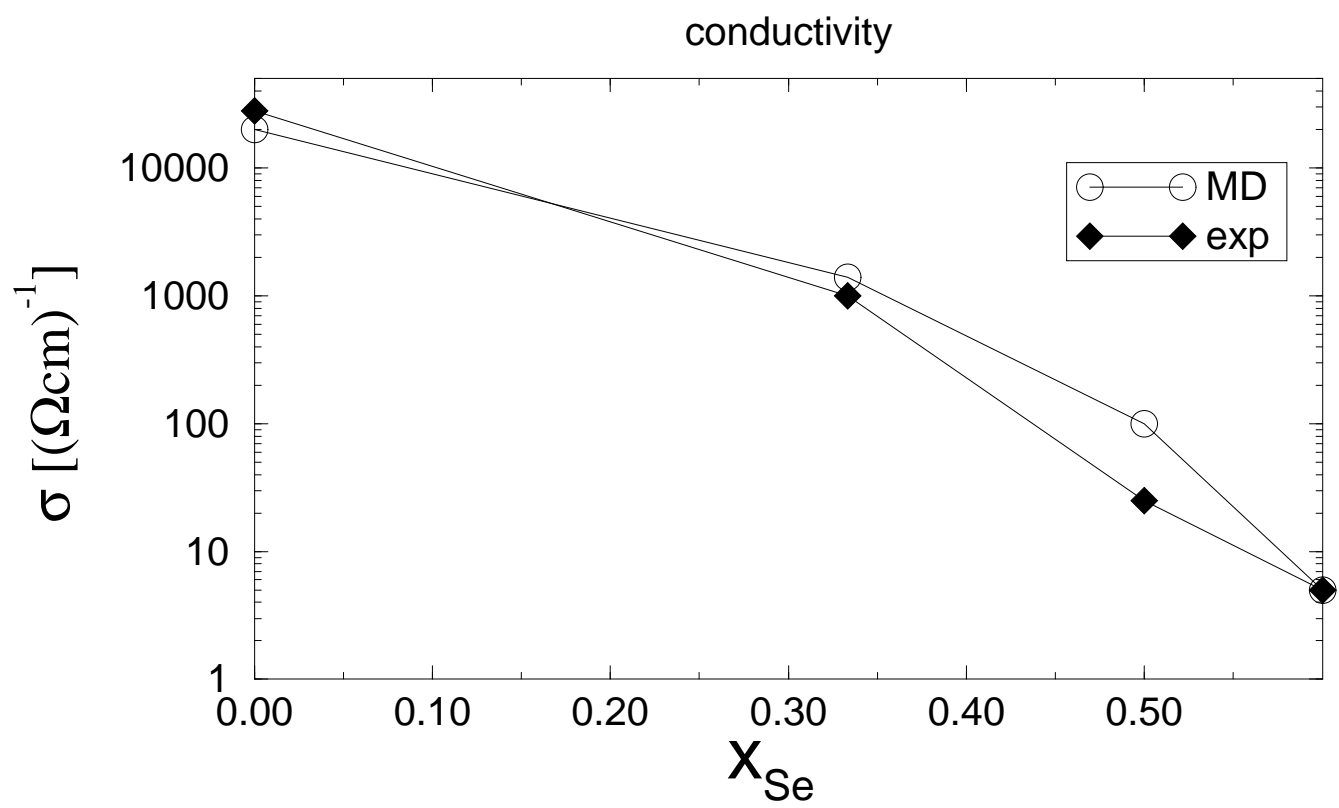
\end{document}